\documentclass[
reprint,
superscriptaddress,
amsmath,amssymb,
aps,
longbibliography,
prl,
]{revtex4-2}

\usepackage{amsbsy,amssymb,amsmath,amsthm,mathtools}
\usepackage{xcolor}
\usepackage{graphicx}
\usepackage{mathrsfs}

\usepackage{tikz}
\usetikzlibrary{decorations.pathmorphing}

\makeatletter
\PassOptionsToPackage{caption=false}{subfig} 
\usepackage{hyperref}
\hypersetup{
breaklinks=true,
colorlinks=true,
citecolor=blue,
linkcolor=black,
filecolor=black,
urlcolor=blue
}
\IfFileExists{lmodern.sty}{\usepackage{lmodern}}{}

\def\be{\begin{equation}}
\def\ee{\end{equation}}

\DeclareMathOperator{\Id}{Id}

\begin{document}

\title{Locality in Dissipative Ground-State Preparation}

\author{Qicheng Tang}
\affiliation{School of Physics, Georgia Institute of Technology, Atlanta, GA 30332, USA}

\author{Ruhanshi Barad}
\affiliation{School of Physics, Georgia Institute of Technology, Atlanta, GA 30332, USA}

\author{Xueda Wen}
\affiliation{School of Physics, Georgia Institute of Technology, Atlanta, GA 30332, USA}

\begin{abstract}
Can the ground state of a local Hamiltonian be prepared by local dissipation? 
We address this question for free-fermion Hamiltonians, using a quadratic Lindbladian with linear jump operators for the dissipative ground-state preparation. 
For translation-invariant targets, Fourier correspondence relates spatial locality to momentum-space analyticity, allowing us to derive a complete spectral criterion for the existence of exponentially local linear jump operators that prepare the ground state. 
This criterion leads to a no-go theorem that, sign-changing Fermi-level crossings exclude exponentially local linear jump operators for preparing the ground state. 
While these crossings are the complete obstruction in one dimension, additional obstructions generally arise in higher dimensions. 
We illustrate these additional obstructions using two quadratic-band-touching models with algebraically decaying ground-state correlations.
We further show that short-range ground-state correlations, although not necessary, are sufficient for local dissipative preparation even without translation invariance. 
Our results establish how the spectral structure of the target Hamiltonian governs the locality of dissipative ground-state preparation.
\end{abstract}
\maketitle

\paragraph{Introduction.---} 
Preparing ground states of quantum many-body systems is a fundamental yet challenging goal of quantum simulation.
Among various approaches, including adiabatic state preparation \cite{lidar_2018_rmp, sipser_2000_adibatic},
imaginary-time evolution \cite{Motta_2019,McArdle_2019},
and coherent Hamiltonian deformations \cite{2021_Zaletel,Kuzmin_2022,Goto:2021,2019_Mitra,Wen_2022}, engineered dissipation has attracted considerable attention recently \cite{Cirac_2008_dissipation, 2008_Kraus, 2013_Cormick, Ge_2018, Goldstein_2019, 2020_gefen, 2023_cubitt, 2024_Ding, Li_2024, Eder_2025, 2026_Zhan, 2026_Ding, 2024_google_science, 2025_Lloyd, 2026_buchhold_FF, 2026_choi_ff, 2024_langbehn, 2026_jian_adaptive, You_2026,2026_elman_chem_TS, 2026_wu,2026_Tang,2026_Bai}. 
By designing dissipation that removes excitations while preserving the ground-state subspace, engineered dissipation provides a natural route to ground-state preparation, potentially from arbitrary initial states.
Rapid convergence of such dissipative protocols has recently been established for several classes of free and weakly interacting models  \cite{Fang_2024, Li_2024, Lin_2025,  2026_Zhan, 2026_Ding}.  
More recently, dissipative ground-state preparation has been experimentally demonstrated in a one-dimensional transverse-field Ising chain using a trapped-ion quantum simulator \cite{Seki_2026}. Engineered dissipation has also been explored for preparing topologically ordered states, including toric-code and fractional quantum Hall states \cite{Dengis_2014,Koenig_2013,Liu_2021}.

While most existing literature focus on the time complexity of realizing the dissipative ground-state preparation, the spatial structure of these dissipations is comparably less studied \cite{elman_2026_local_filter}. 
On the one hand, removing excitations requires a spectral construction, where jump operators are specified through allowing only energy-lowering transitions between eigenstates of the target Hamiltonian. 
On the other hand, physical realizations would prefer the constructed jump operators to be geometrically local in real space. 
This raises a basic question: \textit{does a local target Hamiltonian necessarily admit local dissipative ground-state preparation, or can its spectral structure require nonlocal jump operators?}

In this letter, we address this question for geometrically local free-fermion Hamiltonians in arbitrary spatial dimensions, with considering a quadratic Lindbladian for dissipative ground-state preparation. 
Throughout, \emph{geometric locality} is not restricted to strictly finite range but also includes quasi-locality with exponentially decaying spatial tails.
For translation-invariant target Hamiltonians, the momentum-space description connects spectral structure with spatial locality, allowing us to establish a necessary and sufficient criterion for local preparation with a dark ground-state subspace.

In particular, we prove a no-go theorem: for a translation-invariant free-fermion Hamiltonian with sign-changing Fermi-level crossings, no local linear jumps can prepare its ground state.
The crossing obstruction is complete in one dimension, whereas in higher dimensions the absence of sign-changing crossings does not guarantee local preparation.
We illustrate this distinction with two quadratic-band-touching models in two dimensions: both have algebraically decaying ground-state correlations, but only one admits local linear jumps for ground state preparation.  
While short-range correlations are not necessary for local preparation, we show that they are sufficient even without translation invariance.
For any geometrically local free-fermion Hamiltonian with exponentially decaying ground-state correlations, we construct a class of local jump operators that prepare the ground state.

\paragraph{Dissipative ground-state preparation.---}
Let $\mathbf H$ be the target Hamiltonian, and consider the Lindbladian dynamics
\be
\frac{\mathrm d\rho}{\mathrm dt}
=\mathcal L(\rho)
=-i[\mathbf H,\rho]
+\sum_\mu \mathcal{D}[K_\mu](\rho) ,
\ee
with 
$\mathcal{D}[K_\mu](\rho) = K_\mu\rho K_\mu^\dagger -\tfrac12\{K_\mu^\dagger K_\mu,\rho\}$. 
Here $\rho$ is the density matrix of system under evolution, $\mu$ labels dissipative channels and $K_\mu$ denotes the associated jump operators. 
The aim is to relax the system to the target ground-state subspace $\Omega_{\mathbf H}$ of $\mathbf H$, with $P_{\rm GS}$ denoting the projector onto it. 
While the coherent evolution leaves every eigenstate of the target Hamiltonian unchanged, we further impose the \emph{dark-state condition} 
\be
K_\mu P_{\rm GS}=0\qquad\text{for every }\mu,
\label{eq:dark-state-condition}
\ee
so that every density matrix supported on $\Omega_{\mathbf H}$ is stationary under the full dynamics. 

Preparation from arbitrary initial state additionally requires global attraction of the target ground-state subspace $\Omega_{\mathbf H}$, i.e. 
\be
\lim_{t\to\infty}
\operatorname{Tr}\!\left[(I-P_{\rm GS})e^{t\mathcal L}\rho_0\right]=0
\qquad\text{for every }\rho_0.
\label{eq:global-attraction-condition}
\ee
Together with the darkness condition in Eq.~\eqref{eq:dark-state-condition}, Eq.~\eqref{eq:global-attraction-condition} ensures the convergence of the Lindbladian dynamics to a stationary state supported on $\Omega_{\mathbf H}$. 
If the target Hamiltonian $\mathbf H$ has a non-degenerate ground state $|\mathrm{GS}\rangle$, then $\mathcal{L}$ has a unique steady state $\rho_\infty = \rho(t \to \infty) =|\mathrm{GS}\rangle\langle\mathrm{GS}|$. 
While for degenerate ground states, there could be multiple steady states, and the stationary state $\rho_\infty$ may depend on the initial condition $\rho_0$.

\paragraph{Free-fermion models and linear jumps.---}
Consider a general free-fermion model on a $d$-dimensional lattice with fermionic creation (annihilation) operators $c^\dagger_j$ ($c_j$), $j=1,\dots,N$ defined on the $N$ sites of the lattice. In the Nambu basis  $\Psi=(c_1,\ldots,c_N,c_1^\dagger,\ldots,c_N^\dagger)^T$, with $\Psi_{j,1}=c_j$ and $\Psi_{j,2}=c_j^\dagger$, the Hamiltonian has the standard Bogoliubov--de~Gennes (BdG) form
\be
\mathbf H=\tfrac12\Psi^\dagger h\Psi ,
\ee
upto an additive constant. Here $h$ is the $2N \times 2N$ BdG kernel. We assume that the couplings are geometrically local with possible exponential tails, with uniform bounds on the coupling strength and the scale of spatial decay across spatial positions and system sizes. This leads to a spectral bound $\| h \| \le E_{\rm max}$, where $E_{\rm max} = \mathcal O(1) > 0$ is a size-independent constant. 

Let $h \psi_\nu = \varepsilon_\nu \psi_\nu$ be the eigenvalue decomposition of $h$, with single-particle energies $\varepsilon_\nu$ and eigenvectors $\psi_\nu$ ($\nu = 1,\dots,2N$). 
The mode operators $\gamma_\nu=\psi_\nu^\dagger\Psi$ satisfy $[\mathbf H,\gamma_\nu]=-\varepsilon_\nu\gamma_\nu$, so that the positive-energy modes ($\varepsilon_\nu>0$) lower the many-body energy and annihilate the ground-state subspace as $\gamma_\nu P_{\rm GS}=0$. A linear jump operator $K_\mu=\sum_{j,a}G_{\mu;(j,a)}\Psi_{j,a}$ satisfies the darkness condition in Eq.~\eqref{eq:dark-state-condition} if and only if 
\be
K_\mu
=\sum_{\nu:\varepsilon_\nu>0}
\ell_{\mu\nu}\gamma_\nu, \quad
G_{\mu;(j,a)}
=\sum_{\nu:\varepsilon_\nu>0}
\ell_{\mu\nu}[\psi_\nu]^*_{j,a},
\label{eq:linear-dark-jump}
\ee
i.e. contains only positive-energy modes. Equaivalently, the coefficient row $G_\mu$ annihilates every eigenvector with non-positive energy, as $G_\mu\psi_\nu=0$ for $\varepsilon_\nu\leq0$. 

Moreover, satisfying the global-attraction condition in Eq.~\eqref{eq:global-attraction-condition} requires each positive-energy mode to be coupled to at least one dissipative channel, giving $\sum_\mu|G_\mu\psi_\nu|^2>0$ for $\varepsilon_\nu>0$. 
Together, these state-preparation requirements impose a spectral selection~\footnote{It is in general not a projection, since every channel can couple to each mode with different decay rates.}: all components with non-positive energy must be untouched, while those with positive energy must be coupled to dissipation.

\paragraph{Spatial locality and spectral selection.---}
The coefficients $G_{\mu;(j,a)}$ in Eq.~\eqref{eq:linear-dark-jump} determine the spatial structure of the linear jumps for ground-state preparation. 
For a channel $\mu$ centered at spatial position $x_\mu$, exponential locality requires those coefficients decay exponentially with uniform bounds across channels and system sizes. The question is whether such locality requirements are compatible with the spectral selection discussed above. 

To make this connection, we consider a translation-invariant target Hamiltonian with $s$ orbitals per cell. Periodic boundary conditions are imposed for explicit translation invariance. 
Write $j=(x,\alpha)$, with $x \in\mathbb Z^d$ labels cells and $\alpha=1,\dots,s$ labels orbitals, and correspondingly $c_j=c_{x,\alpha}$. 
Due to the translation invariance, the hopping and pairing amplitudes can be collected into blocks $h_r$, of size $2s\times2s$, that depend only on the displacement $r$ between cells.
On the Brillouin zone (BZ), the Fourier transformation of these blocks gives the Bloch kernels
\be
h(q)=\sum_{r\in\mathbb Z^d}e^{iq\cdot r}h_r,
\qquad q\in{\rm BZ}=[-\pi,\pi)^d.
\ee
Denote their eigenvalue decompositions by $h(q)u_n(q)=\varepsilon_n(q)u_n(q)$, with band energies $\varepsilon_n(q)$ and Bloch eigenvectors $u_n(q)$ labeled by the band index $n=1,\ldots,2s$. 
At a finite system size $L$, the momentum $q$ takes the allowed values $q_k = 2\pi k/L$ within the BZ, with $k\in \mathbb Z^d$. 

The Fourier transform relates spatial locality to analyticity in momentum: a continuous periodic function has exponentially decaying Fourier coefficients if and only if it admits an analytic extension to a complex neighborhood of its real domain.
For the linear jumps in Eq.~\eqref{eq:linear-dark-jump}, we define the Fourier transformation of their real-space coefficients 
\be
[G_\mu(q)]_{(\alpha,a)}
=
\sum_{x\in\mathbb Z^d}e^{-iq\cdot x}
G_{\mu;((x,\alpha),a)}.
\label{eq:jump-coefficient-Fourier}
\ee
Applying the Fourier correspondence, geometric locality of the jump operators is then equivalent to each $G_\mu(q)$ being analytic on a complex neighborhood of the BZ.
Here, we use the same functions $G_\mu(q)$ across all system sizes $L$, sampled at the allowed momenta $q_k=2\pi k/L$ for each finite size. 

In particular, at each allowed momentum $q_k$ on finite lattice, the preparation requirements become
\begin{align}
G_\mu(q_k)u_n(q_k)&=0
&&\text{if }\varepsilon_n(q_k)\leq0,
\label{eq:momentum-dark-condition}\\
\sum_\mu|G_\mu(q_k)u_n(q_k)|^2&>0
&&\text{if }\varepsilon_n(q_k)>0.
\label{eq:momentum-global-attraction}
\end{align}
Here, the first line is the dark-state condition. The second line corresponds to the requirement for global attraction, and it is generally a necessary condition since we do not require translation invariance of the jump operators associated with different channels. 

\paragraph{A locality criterion for target spectra.---}
To see how these conditions along with analyticity of $G_\mu(q)$ relate to the spectral structure of the target Hamiltonian, let us start with the simplest case of one orbital per cell, i.e. a single-band model, with particle-hole symmetry. The $2\times2$ Bloch kernel $h(q)$ then has eigenvalues $\pm E(q)$, $E(q) = |\varepsilon(q)| \ge 0$. For $E(q) > 0$, Eq.~\eqref{eq:momentum-dark-condition} gives
\be
G_\mu(q) h(q) = E(q) G_\mu(q) ,
\ee
and Eq.~\eqref{eq:momentum-global-attraction} ensures that there is a component $[G_\mu(q)]_a$ non-vanishing for some channels, giving 
\be
E(q)=\frac{[G_\mu(q)h(q)]_a}{[G_\mu(q)]_a}.
\ee
Assuming $G_\mu(q)$ is analytic for making the jump operator local in real space, then the above equation tells us $E(q)$ can be expressed as the quotient of two analytic functions (here $h(q)$ is analytic followed by the real-space locality of $h$), and therefore should be meromorphic. 
For this single-band model, this is to require that the positive-energy branch of the spectra cannot crosses the Fermi-level, therefore providing a restriction to the spectral structure of the target Hamiltonian for a local ground-state preparation. 

The above arguments can be extended to general multi-band cases. Instead of requiring each band to have no Fermi-level crossings, it has to be considered in a matrix sense, as a single row $G_\mu(q)$ could couple to several different positive-energy modes at the same momentum.
We therefore consider a matrix $M(q)$ that is consist of the powers $G_\mu(q) [h(q)]^k$, with $k = 0, 1, \dots, 2s-1$, as~\footnote{Here $h(q)$ is of rank $2s$, so that any power higher than $2s-1$ can be expressed in terms of lower powers $k=0,1,\dots,2s-1$.}
\be
M(q) = \begin{pmatrix}
    G_{\mu_1} h(q)^{k_1} \\ \vdots \\ G_{\mu_r} h(q)^{k_r} 
\end{pmatrix}.
\ee
Here $r$ is the maximum number of independent rows in such construction, and equal to the number of positive-energy modes at momentum $q$. 
$M(q)$ is therefore a matrix of size $r \times 2s$ that gives
\be
M(q) h(q) = B(q) M(q).
\ee
Suppose $A$ is a set of $r$ columns for which $[M(q)]_A$ is invertible
, then one has
\be
B(q) = [M(q) h(q)]_A \{ [M(q)]_A \}^{-1},
\ee
whose eigenvalues are the positive energies of $h(q)$. Again, to let the linear jumps for ground-state preparation exponentially local in real space, $B(q)$ must be a meromorphic matrix function. 

While the above construction already can lead to obstructions, the obtained $B(q)$ would depend on the choice of $M(q)$, with a difference upto a similarity transformation on $B(q) \to \widetilde B(q) = S B(q) S^{-1}$. It therefore motivates us to consider 
\be
\chi_+(z, q) = \det\left[ z \cdot \Id - B(q) \right],
\label{eq:positive-character}
\ee
which is generally meromorphic in $q$. 
Moreover, under a suitable choice of the coordinate (rotated by similarity transformation applied onto $h(q)$), the positive- and non-positive-energy branches of $h(q)$ can be in a block form, leading to 
\be
\det\left[ z \cdot \Id - h(q) \right] = \chi_+(z, q) \chi_-(z, q)
\label{eq:locality-criterion}
\ee
where $\chi_+(z, q)$ and $\chi_-(z, q)$ denote the characteristic polynomials for positive and non-positive energies, respectively. 
Here, since $h(q)$ is bounded, the coefficients in $\chi_\pm(z, q)$ are bounded and imply the possible meromorphic singularities in them are removable. Thus, Eq.~\eqref{eq:locality-criterion} establishes an analytic factorization for exponentially local jump operators. 

Conversely, Eq.~\eqref{eq:locality-criterion} also provides an explicit construction for local jumps. Choose the $2s\times2s$ coefficient matrix
\be
G(q) = \sqrt{\kappa} \, h(q) \, \chi_-(h(q),q)
\label{eq:spectra-construct-local-jump}
\ee
where $\kappa>0$ sets the dissipative rate scale.
Its rows and their lattice translates define $2s$ channels per cell, labeled by $\mu=((x,\beta),b)$, with Fourier coefficients
\be
[G_\mu(q)]_{(\alpha,a)}
=
e^{-iq\cdot x}[G(q)]_{(\beta,b),(\alpha,a)}.
\ee
Here $\beta=1,\ldots,s$ and $b=1,2$ are the orbital and Nambu indices, respectively.

Since $h(q)$ and $\chi_-$ are both analytic, $G(q)$ is analytic as well. Meanwhile, it acts onto a Bloch eigenvector as
\be
G(q)u_n(q)
= \sqrt{\kappa}\, 
\varepsilon_n(q)\, \chi_-(\varepsilon_n(q),q) u_n(q).
\label{eq:constructed-spectral-action}
\ee
By definition, $\chi_-(\varepsilon_n(q),q)=0$ for $\varepsilon_n(q)\le0$ and $\chi_-(\varepsilon_n(q),q)>0$ for $\varepsilon_n(q)>0$. In summary, the construction in Eq.~\eqref{eq:spectra-construct-local-jump} leads to local jumps for ground-state preparation. 
Consequently, the analytic factorization in Eq.~\eqref{eq:locality-criterion} establishes a \emph{necessary and sufficient criterion} for exponential locality of linear jump operators. 

\paragraph{No-go for Fermi-level crossings.---}
Consider a real analytic path $q(\lambda)$ along which an analytic eigenvalue function $\varepsilon_n(\lambda)$ crosses the Fermi level with a sign change at $\lambda=0$.~\footnote{The existence of such choice is ensured by the Rellich's theorem~\cite{Rellich1937, Kato1995}. It states that along a real-analytic path $q(\lambda)$ in the BZ, the Hermitian matrix $h[q(\lambda)]$ (which is analytic in $q$ followed by the real-space locality of $h$) allows to choose its eigenvalues $\varepsilon_n(\lambda)$ and corresponding eigenvectors $u_n(\lambda)$ as analytic functions of $\lambda$.} 
Assume the orientation of $q(\lambda)$ is chosen such that  $\varepsilon_n(\lambda)<0$ for $\lambda<0$ and $\varepsilon_n(\lambda)>0$ for $\lambda>0$ near the crossing.
Since every negative energy is a root of $\chi_-$,
\be
\chi_-\!\left(\varepsilon_n(\lambda),q(\lambda)\right)=0
\qquad(\lambda<0).
\ee
The left-hand side is analytic in $\lambda$ and therefore vanishes identically near the crossing.
For $\lambda>0$, this would make a positive energy a root of $\chi_-$, contradicting Eq.~\eqref{eq:locality-criterion}.
Thus, we arrive at a \emph{no-go theorem}: no geometrically linear jump operator is allowed for preparing the ground state for a target Hamiltonian with sign-changing Fermi-level crossing. 

It is known that a Hamiltonian with Fermi-level crossing has algebraically decaying ground-state correlations, which is the signature of criticality in the system. It is therefore natural to ask whether every critical ground state cannot be prepared locally with linear jumps. 
In one dimension, the eigenvectors and eigenvalues of $h(q)$ are locally analytic in $q$, and an analytic energy branch touches zero without changing sign remains either nonnegative or nonpositive on both sides of the touching. Consequently, $|h(q)|=[h(q)^2]^{1/2}$ is analytic, and a simple choice
\be
G(q)\propto h(q)P_+(q)
=\frac12\left[h(q)+|h(q)|\right] 
\ee
gives exponentially local jumps for ground-state preparation. Here $P_+(q)$ denotes the projector to positive-energy bands at momentum $q$. Therefore, sign-changing Fermi-level crossings are the only obstruction in one dimension. 

\emph{Obstruction for quadratic band touching.---}
In contrast, higher dimensions allow additional obstructions.
We illustrate this with two models that both exhibit quadratic
band touching and algebraically decaying ground-state correlations,
but only one admits ground-state preparation by local linear jumps.
For the first model, its hopping matrix is given by 
\be
A(q)=(\sin^2q_x-\sin^2q_y)\sigma_z
+2\sin q_x\sin q_y\,\sigma_x,
\label{eq:QBT-local-jump}
\ee
where $\sigma_{i}$, $i=x,y,z$ denotes Pauli operators. 
Its bands have energies $\pm E(q)$, with $E(q)=\sin^2q_x+\sin^2q_y$, and the BdG kernel is $h(q)=\operatorname{diag}[A(q),-A(q)]$.
For this model, the coefficient kernel
\be
G(q)\propto h(q)P_+(q)
=\frac12[h(q)+E(q)\Id]
\ee
is a trigonometric polynomial, and gives strictly finite-range jumps for preparing its ground state. 
For comparison, let us consider the second model, 
\be
A'(q)=(1-\cos q_x)\sigma_z+(1-\cos q_y)\sigma_x,
\ee
with $h'(q)=\operatorname{diag}[A'(q),-A'(q)]$.
Its bands also touch quadratically without changing sign, but the positive energy
$ E'(q)=\sqrt{(1-\cos q_x)^2+(1-\cos q_y)^2} $
is nonanalytic at the origin. 
The required spectral factors $\chi_\pm(z,q)=[z\mp E'(q)]^2$ therefore contain the nonanalytic coefficients $\mp2E'(q)$, violating the locality criterion established in Eq.~\eqref{eq:locality-criterion}.
Thus, despite their similar low-energy dispersions and algebraically decaying correlations, the two models exhibit fundamentally different locality properties for dissipative ground-state preparation.

\paragraph{Local preparation for short-range correlations.---}
The example in Eq.~\eqref{eq:QBT-local-jump} shows that short-range ground-state correlations are not necessary for local preparation.
We now show that they are sufficient, even without translation invariance.
Suppose the target Hamiltonian $h$ admits a ground state whose Nambu correlation matrix 
$C_{(i,a),(j,b)}=\langle\Psi_{j,b}^\dagger\Psi_{i,a}\rangle_{\rm GS}$ decays exponentially with distance between sites $i$ and $j$. 
Denote $P_+$ as the projector onto the positive-energy subspace of $h$, and introduce the real-space coefficient kernel 
\be
G \propto hP_+ =h(\Id-C)
\label{eq:linear-weight-jump-coefficient}
\ee
that by design prepares the target ground state. 
As both $h$ and $\Id-C$ have exponentially decaying matrix elements, the constructed kernel $G$ and corresponding jump operators are geometrically local. 
Therefore, \emph{any free-fermion system with short-range ground-state correlations can be prepared using a local Lindbladian}. 

\paragraph{Summary and discussion.---}
In this Letter, we address the question about geometric locality of linear jump operators for ground-state preparation through a Lindbladian dynamics. With translation invariance, we derive a criterion that relates the locality of jump operators to the spectral structure of target Hamiltonian. In particular, this locality criterion leads to a no-go for locally preparing the ground state of a general $d$-dimensional free-fermion Hamiltonian with sign-changing Fermi-level crossing in its band structure. In one dimension, the Fermi-level crossing is the only obstruction, but additional obstructions could occur in higher dimensions. 

It is natural to consider the relation between the locality of jump operators and ground-state correlations of target Hamiltonian. A simple setting for the jump operators is to choose $G \propto P_+ = \Id - C$, i.e. the projector onto the positive-energy modes. One may then consider algebraically decaying ground-state correlations require non-local jump operators. However, the projector design is only a single choice among infinite constructions. In particular, we provide two examples with both quadratic band touching and algebraically decaying ground-state correlations, but only one admits ground-state preparation with local linear jumps. 

Nevertheless, for target Hamiltonian with short-range ground-state correlations, we show that a simple design of the jump operator, shown in Eq.~\eqref{eq:linear-weight-jump-coefficient}, can produce local preparation, without requiring translation invariance. Moreover, such jumps can be realized by a spectral-filtering approach that allows a fast mixing towards to the ground state and an efficient evaluation without knowing spectral information of the target Hamiltonian, as demonstrated in the End Matter. 

It should be emphasized that, the compatibility between local dissipation and algebraic correlations in the corresponding steady state has been studied previously~\cite{2022_Zhang}. In particular, Ref.~\cite{2022_Zhang} considered a finite-range translation-invariant quadratic Lindbladian, with assuming a unique steady state. In such setting, it is found that, the steady state correlations cannot exhibit algebraic decay for free fermions in any spatial dimensions. Here, by allowing multiple steady state (while requires the steadiness from a given initial state) and exponential detail of linear jump operators, we are able to locally prepare a target ground state with algebraic correlations in spatial dimensions higher than one. 

Another direction for future work is to extend our analysis to interacting quantum many-body systems, such as general one-dimensional spin chains. It is interesting to determine whether analogous obstructions to local dissipative ground-state preparation arise in interacting critical systems. Unlike free-fermion systems, where the locality of jump operators can be characterized through the analytic structure of single-particle spectral kernels, interacting systems require a more general understanding of the relationship between many-body spectra and operator locality.

\paragraph{Acknowledgments.}
This work is supported by a startup at Georgia Institute of Technology. 
The authors used AI for language editing and for refining mathematical arguments. 

\textit{Note added}:
During the completion of this draft, we become aware of a related work \cite{2026_Westhoff} appeared on arXiv. 

\bibliographystyle{apsrev4-2}
\bibliography{local_jump}

\begin{thebibliography}{46}%
\makeatletter
\providecommand \@ifxundefined [1]{%
 \@ifx{#1\undefined}
}%
\providecommand \@ifnum [1]{%
 \ifnum #1\expandafter \@firstoftwo
 \else \expandafter \@secondoftwo
 \fi
}%
\providecommand \@ifx [1]{%
 \ifx #1\expandafter \@firstoftwo
 \else \expandafter \@secondoftwo
 \fi
}%
\providecommand \natexlab [1]{#1}%
\providecommand \enquote  [1]{``#1''}%
\providecommand \bibnamefont  [1]{#1}%
\providecommand \bibfnamefont [1]{#1}%
\providecommand \citenamefont [1]{#1}%
\providecommand \href@noop [0]{\@secondoftwo}%
\providecommand \href [0]{\begingroup \@sanitize@url \@href}%
\providecommand \@href[1]{\@@startlink{#1}\@@href}%
\providecommand \@@href[1]{\endgroup#1\@@endlink}%
\providecommand \@sanitize@url [0]{\catcode `\\12\catcode `\$12\catcode
  `\&12\catcode `\#12\catcode `\^12\catcode `\_12\catcode `\%12\relax}%
\providecommand \@@startlink[1]{}%
\providecommand \@@endlink[0]{}%
\providecommand \url  [0]{\begingroup\@sanitize@url \@url }%
\providecommand \@url [1]{\endgroup\@href {#1}{\urlprefix }}%
\providecommand \urlprefix  [0]{URL }%
\providecommand \Eprint [0]{\href }%
\providecommand \doibase [0]{https://doi.org/}%
\providecommand \selectlanguage [0]{\@gobble}%
\providecommand \bibinfo  [0]{\@secondoftwo}%
\providecommand \bibfield  [0]{\@secondoftwo}%
\providecommand \translation [1]{[#1]}%
\providecommand \BibitemOpen [0]{}%
\providecommand \bibitemStop [0]{}%
\providecommand \bibitemNoStop [0]{.\EOS\space}%
\providecommand \EOS [0]{\spacefactor3000\relax}%
\providecommand \BibitemShut  [1]{\csname bibitem#1\endcsname}%
\let\auto@bib@innerbib\@empty
\bibitem [{\citenamefont {{Albash}}\ and\ \citenamefont
  {{Lidar}}(2018)}]{lidar_2018_rmp}%
  \BibitemOpen
  \bibfield  {author} {\bibinfo {author} {\bibfnamefont {T.}~\bibnamefont
  {{Albash}}}\ and\ \bibinfo {author} {\bibfnamefont {D.~A.}\ \bibnamefont
  {{Lidar}}},\ }\href {https://doi.org/10.1103/RevModPhys.90.015002} {\bibfield
   {journal} {\bibinfo  {journal} {Reviews of Modern Physics}\ }\textbf
  {\bibinfo {volume} {90}},\ \bibinfo {eid} {015002} (\bibinfo {year}
  {2018})},\ \Eprint {https://arxiv.org/abs/1611.04471} {arXiv:1611.04471
  [quant-ph]} \BibitemShut {NoStop}%
\bibitem [{\citenamefont {{Farhi}}\ \emph {et~al.}(2000)\citenamefont
  {{Farhi}}, \citenamefont {{Goldstone}}, \citenamefont {{Gutmann}},\ and\
  \citenamefont {{Sipser}}}]{sipser_2000_adibatic}%
  \BibitemOpen
  \bibfield  {author} {\bibinfo {author} {\bibfnamefont {E.}~\bibnamefont
  {{Farhi}}}, \bibinfo {author} {\bibfnamefont {J.}~\bibnamefont
  {{Goldstone}}}, \bibinfo {author} {\bibfnamefont {S.}~\bibnamefont
  {{Gutmann}}},\ and\ \bibinfo {author} {\bibfnamefont {M.}~\bibnamefont
  {{Sipser}}},\ }\href {https://doi.org/10.48550/arXiv.quant-ph/0001106}
  {\bibfield  {journal} {\bibinfo  {journal} {arXiv e-prints}\ ,\ \bibinfo
  {eid} {quant-ph/0001106}} (\bibinfo {year} {2000})},\ \Eprint
  {https://arxiv.org/abs/quant-ph/0001106} {arXiv:quant-ph/0001106 [quant-ph]}
  \BibitemShut {NoStop}%
\bibitem [{\citenamefont {{Motta}}\ \emph {et~al.}(2020)\citenamefont
  {{Motta}}, \citenamefont {{Sun}}, \citenamefont {{Tan}}, \citenamefont
  {{O'Rourke}}, \citenamefont {{Ye}}, \citenamefont {{Minnich}}, \citenamefont
  {{Brand{\~a}o}},\ and\ \citenamefont {{Chan}}}]{Motta_2019}%
  \BibitemOpen
  \bibfield  {author} {\bibinfo {author} {\bibfnamefont {M.}~\bibnamefont
  {{Motta}}}, \bibinfo {author} {\bibfnamefont {C.}~\bibnamefont {{Sun}}},
  \bibinfo {author} {\bibfnamefont {A.~T.~K.}\ \bibnamefont {{Tan}}}, \bibinfo
  {author} {\bibfnamefont {M.~J.}\ \bibnamefont {{O'Rourke}}}, \bibinfo
  {author} {\bibfnamefont {E.}~\bibnamefont {{Ye}}}, \bibinfo {author}
  {\bibfnamefont {A.~J.}\ \bibnamefont {{Minnich}}}, \bibinfo {author}
  {\bibfnamefont {F.~G.~S.~L.}\ \bibnamefont {{Brand{\~a}o}}},\ and\ \bibinfo
  {author} {\bibfnamefont {G.~K.-L.}\ \bibnamefont {{Chan}}},\ }\href
  {https://doi.org/10.1038/s41567-019-0704-4} {\bibfield  {journal} {\bibinfo
  {journal} {Nature Physics}\ }\textbf {\bibinfo {volume} {16}},\ \bibinfo
  {pages} {205} (\bibinfo {year} {2020})},\ \Eprint
  {https://arxiv.org/abs/1901.07653} {arXiv:1901.07653 [quant-ph]} \BibitemShut
  {NoStop}%
\bibitem [{\citenamefont {{McArdle}}\ \emph {et~al.}(2019)\citenamefont
  {{McArdle}}, \citenamefont {{Jones}}, \citenamefont {{Endo}}, \citenamefont
  {{Li}}, \citenamefont {{Benjamin}},\ and\ \citenamefont
  {{Yuan}}}]{McArdle_2019}%
  \BibitemOpen
  \bibfield  {author} {\bibinfo {author} {\bibfnamefont {S.}~\bibnamefont
  {{McArdle}}}, \bibinfo {author} {\bibfnamefont {T.}~\bibnamefont {{Jones}}},
  \bibinfo {author} {\bibfnamefont {S.}~\bibnamefont {{Endo}}}, \bibinfo
  {author} {\bibfnamefont {Y.}~\bibnamefont {{Li}}}, \bibinfo {author}
  {\bibfnamefont {S.~C.}\ \bibnamefont {{Benjamin}}},\ and\ \bibinfo {author}
  {\bibfnamefont {X.}~\bibnamefont {{Yuan}}},\ }\href
  {https://doi.org/10.1038/s41534-019-0187-2} {\bibfield  {journal} {\bibinfo
  {journal} {npj Quantum Information}\ }\textbf {\bibinfo {volume} {5}},\
  \bibinfo {eid} {75} (\bibinfo {year} {2019})},\ \Eprint
  {https://arxiv.org/abs/1804.03023} {arXiv:1804.03023 [quant-ph]} \BibitemShut
  {NoStop}%
\bibitem [{\citenamefont {{Zaletel}}\ \emph {et~al.}(2016)\citenamefont
  {{Zaletel}}, \citenamefont {{Kaufman}}, \citenamefont {{Stamper-Kurn}},\ and\
  \citenamefont {{Yao}}}]{2021_Zaletel}%
  \BibitemOpen
  \bibfield  {author} {\bibinfo {author} {\bibfnamefont {M.~P.}\ \bibnamefont
  {{Zaletel}}}, \bibinfo {author} {\bibfnamefont {A.~M.}\ \bibnamefont
  {{Kaufman}}}, \bibinfo {author} {\bibfnamefont {D.~M.}\ \bibnamefont
  {{Stamper-Kurn}}},\ and\ \bibinfo {author} {\bibfnamefont {N.~Y.}\
  \bibnamefont {{Yao}}},\ }\href {https://doi.org/10.48550/arXiv.1611.04591}
  {\bibfield  {journal} {\bibinfo  {journal} {arXiv e-prints}\ ,\ \bibinfo
  {eid} {arXiv:1611.04591}} (\bibinfo {year} {2016})},\ \Eprint
  {https://arxiv.org/abs/1611.04591} {arXiv:1611.04591 [cond-mat.quant-gas]}
  \BibitemShut {NoStop}%
\bibitem [{\citenamefont {{Kuzmin}}\ \emph {et~al.}(2022)\citenamefont
  {{Kuzmin}}, \citenamefont {{Zache}}, \citenamefont {{Kokail}}, \citenamefont
  {{Pastori}}, \citenamefont {{Celi}}, \citenamefont {{Baranov}},\ and\
  \citenamefont {{Zoller}}}]{Kuzmin_2022}%
  \BibitemOpen
  \bibfield  {author} {\bibinfo {author} {\bibfnamefont {V.}~\bibnamefont
  {{Kuzmin}}}, \bibinfo {author} {\bibfnamefont {T.~V.}\ \bibnamefont
  {{Zache}}}, \bibinfo {author} {\bibfnamefont {C.}~\bibnamefont {{Kokail}}},
  \bibinfo {author} {\bibfnamefont {L.}~\bibnamefont {{Pastori}}}, \bibinfo
  {author} {\bibfnamefont {A.}~\bibnamefont {{Celi}}}, \bibinfo {author}
  {\bibfnamefont {M.}~\bibnamefont {{Baranov}}},\ and\ \bibinfo {author}
  {\bibfnamefont {P.}~\bibnamefont {{Zoller}}},\ }\href
  {https://doi.org/10.1103/PRXQuantum.3.020304} {\bibfield  {journal} {\bibinfo
   {journal} {PRX Quantum}\ }\textbf {\bibinfo {volume} {3}},\ \bibinfo {eid}
  {020304} (\bibinfo {year} {2022})},\ \Eprint
  {https://arxiv.org/abs/2108.12378} {arXiv:2108.12378 [quant-ph]} \BibitemShut
  {NoStop}%
\bibitem [{\citenamefont {Goto}\ \emph {et~al.}(2025)\citenamefont {Goto},
  \citenamefont {Nozaki}, \citenamefont {Ryu}, \citenamefont {Tamaoka},\ and\
  \citenamefont {Tan}}]{Goto:2021}%
  \BibitemOpen
  \bibfield  {author} {\bibinfo {author} {\bibfnamefont {K.}~\bibnamefont
  {Goto}}, \bibinfo {author} {\bibfnamefont {M.}~\bibnamefont {Nozaki}},
  \bibinfo {author} {\bibfnamefont {S.}~\bibnamefont {Ryu}}, \bibinfo {author}
  {\bibfnamefont {K.}~\bibnamefont {Tamaoka}},\ and\ \bibinfo {author}
  {\bibfnamefont {M.~T.}\ \bibnamefont {Tan}},\ }\href
  {https://doi.org/10.1007/JHEP08(2025)186} {\bibfield  {journal} {\bibinfo
  {journal} {JHEP}\ }\textbf {\bibinfo {volume} {08}},\ \bibinfo {pages}
  {186}},\ \Eprint {https://arxiv.org/abs/2112.14388} {arXiv:2112.14388
  [hep-th]} \BibitemShut {NoStop}%
\bibitem [{\citenamefont {{Mitra}}\ \emph {et~al.}(2019)\citenamefont
  {{Mitra}}, \citenamefont {{Ippoliti}}, \citenamefont {{Bhatt}}, \citenamefont
  {{Sondhi}},\ and\ \citenamefont {{Agarwal}}}]{2019_Mitra}%
  \BibitemOpen
  \bibfield  {author} {\bibinfo {author} {\bibfnamefont {P.}~\bibnamefont
  {{Mitra}}}, \bibinfo {author} {\bibfnamefont {M.}~\bibnamefont {{Ippoliti}}},
  \bibinfo {author} {\bibfnamefont {R.~N.}\ \bibnamefont {{Bhatt}}}, \bibinfo
  {author} {\bibfnamefont {S.~L.}\ \bibnamefont {{Sondhi}}},\ and\ \bibinfo
  {author} {\bibfnamefont {K.}~\bibnamefont {{Agarwal}}},\ }\href
  {https://doi.org/10.1103/PhysRevB.99.104308} {\bibfield  {journal} {\bibinfo
  {journal} {\prb}\ }\textbf {\bibinfo {volume} {99}},\ \bibinfo {eid} {104308}
  (\bibinfo {year} {2019})},\ \Eprint {https://arxiv.org/abs/1809.01681}
  {arXiv:1809.01681 [cond-mat.stat-mech]} \BibitemShut {NoStop}%
\bibitem [{\citenamefont {Wen}\ \emph {et~al.}(2022)\citenamefont {Wen},
  \citenamefont {Fan},\ and\ \citenamefont {Vishwanath}}]{Wen_2022}%
  \BibitemOpen
  \bibfield  {author} {\bibinfo {author} {\bibfnamefont {X.}~\bibnamefont
  {Wen}}, \bibinfo {author} {\bibfnamefont {R.}~\bibnamefont {Fan}},\ and\
  \bibinfo {author} {\bibfnamefont {A.}~\bibnamefont {Vishwanath}},\
  }\href@noop {} {\  (\bibinfo {year} {2022})},\ \Eprint
  {https://arxiv.org/abs/2211.00040} {arXiv:2211.00040 [cond-mat.str-el]}
  \BibitemShut {NoStop}%
\bibitem [{\citenamefont {{Verstraete}}\ \emph {et~al.}(2009)\citenamefont
  {{Verstraete}}, \citenamefont {{Wolf}},\ and\ \citenamefont
  {{Cirac}}}]{Cirac_2008_dissipation}%
  \BibitemOpen
  \bibfield  {author} {\bibinfo {author} {\bibfnamefont {F.}~\bibnamefont
  {{Verstraete}}}, \bibinfo {author} {\bibfnamefont {M.~M.}\ \bibnamefont
  {{Wolf}}},\ and\ \bibinfo {author} {\bibfnamefont {J.~I.}\ \bibnamefont
  {{Cirac}}},\ }\href {https://doi.org/10.1038/nphys1342} {\bibfield  {journal}
  {\bibinfo  {journal} {Nature Physics}\ }\textbf {\bibinfo {volume} {5}},\
  \bibinfo {pages} {633} (\bibinfo {year} {2009})},\ \Eprint
  {https://arxiv.org/abs/0803.1447} {arXiv:0803.1447 [quant-ph]} \BibitemShut
  {NoStop}%
\bibitem [{\citenamefont {{Kraus}}\ \emph {et~al.}(2008)\citenamefont
  {{Kraus}}, \citenamefont {{B{\"u}chler}}, \citenamefont {{Diehl}},
  \citenamefont {{Kantian}}, \citenamefont {{Micheli}},\ and\ \citenamefont
  {{Zoller}}}]{2008_Kraus}%
  \BibitemOpen
  \bibfield  {author} {\bibinfo {author} {\bibfnamefont {B.}~\bibnamefont
  {{Kraus}}}, \bibinfo {author} {\bibfnamefont {H.~P.}\ \bibnamefont
  {{B{\"u}chler}}}, \bibinfo {author} {\bibfnamefont {S.}~\bibnamefont
  {{Diehl}}}, \bibinfo {author} {\bibfnamefont {A.}~\bibnamefont {{Kantian}}},
  \bibinfo {author} {\bibfnamefont {A.}~\bibnamefont {{Micheli}}},\ and\
  \bibinfo {author} {\bibfnamefont {P.}~\bibnamefont {{Zoller}}},\ }\href
  {https://doi.org/10.1103/PhysRevA.78.042307} {\bibfield  {journal} {\bibinfo
  {journal} {\pra}\ }\textbf {\bibinfo {volume} {78}},\ \bibinfo {eid} {042307}
  (\bibinfo {year} {2008})},\ \Eprint {https://arxiv.org/abs/0803.1463}
  {arXiv:0803.1463 [quant-ph]} \BibitemShut {NoStop}%
\bibitem [{\citenamefont {{Cormick}}\ \emph {et~al.}(2013)\citenamefont
  {{Cormick}}, \citenamefont {{Bermudez}}, \citenamefont {{Huelga}},\ and\
  \citenamefont {{Plenio}}}]{2013_Cormick}%
  \BibitemOpen
  \bibfield  {author} {\bibinfo {author} {\bibfnamefont {C.}~\bibnamefont
  {{Cormick}}}, \bibinfo {author} {\bibfnamefont {A.}~\bibnamefont
  {{Bermudez}}}, \bibinfo {author} {\bibfnamefont {S.~F.}\ \bibnamefont
  {{Huelga}}},\ and\ \bibinfo {author} {\bibfnamefont {M.~B.}\ \bibnamefont
  {{Plenio}}},\ }\href {https://doi.org/10.1088/1367-2630/15/7/073027}
  {\bibfield  {journal} {\bibinfo  {journal} {New Journal of Physics}\ }\textbf
  {\bibinfo {volume} {15}},\ \bibinfo {eid} {073027} (\bibinfo {year}
  {2013})},\ \Eprint {https://arxiv.org/abs/1304.2201} {arXiv:1304.2201
  [quant-ph]} \BibitemShut {NoStop}%
\bibitem [{\citenamefont {Ge}\ \emph {et~al.}(2019)\citenamefont {Ge},
  \citenamefont {Tura},\ and\ \citenamefont {Cirac}}]{Ge_2018}%
  \BibitemOpen
  \bibfield  {author} {\bibinfo {author} {\bibfnamefont {Y.}~\bibnamefont
  {Ge}}, \bibinfo {author} {\bibfnamefont {J.}~\bibnamefont {Tura}},\ and\
  \bibinfo {author} {\bibfnamefont {J.~I.}\ \bibnamefont {Cirac}},\ }\href
  {https://doi.org/10.1063/1.5027484} {\bibfield  {journal} {\bibinfo
  {journal} {J. Math. Phys.}\ }\textbf {\bibinfo {volume} {60}},\ \bibinfo
  {pages} {022202} (\bibinfo {year} {2019})},\ \Eprint
  {https://arxiv.org/abs/1712.03193} {arXiv:1712.03193 [quant-ph]} \BibitemShut
  {NoStop}%
\bibitem [{\citenamefont {{Goldstein}}(2019)}]{Goldstein_2019}%
  \BibitemOpen
  \bibfield  {author} {\bibinfo {author} {\bibfnamefont {M.}~\bibnamefont
  {{Goldstein}}},\ }\href {https://doi.org/10.21468/SciPostPhys.7.5.067}
  {\bibfield  {journal} {\bibinfo  {journal} {SciPost Physics}\ }\textbf
  {\bibinfo {volume} {7}},\ \bibinfo {eid} {067} (\bibinfo {year} {2019})},\
  \Eprint {https://arxiv.org/abs/1810.12050} {arXiv:1810.12050
  [cond-mat.quant-gas]} \BibitemShut {NoStop}%
\bibitem [{\citenamefont {{Roy}}\ \emph {et~al.}(2020)\citenamefont {{Roy}},
  \citenamefont {{Chalker}}, \citenamefont {{Gornyi}},\ and\ \citenamefont
  {{Gefen}}}]{2020_gefen}%
  \BibitemOpen
  \bibfield  {author} {\bibinfo {author} {\bibfnamefont {S.}~\bibnamefont
  {{Roy}}}, \bibinfo {author} {\bibfnamefont {J.~T.}\ \bibnamefont
  {{Chalker}}}, \bibinfo {author} {\bibfnamefont {I.~V.}\ \bibnamefont
  {{Gornyi}}},\ and\ \bibinfo {author} {\bibfnamefont {Y.}~\bibnamefont
  {{Gefen}}},\ }\href {https://doi.org/10.1103/PhysRevResearch.2.033347}
  {\bibfield  {journal} {\bibinfo  {journal} {Physical Review Research}\
  }\textbf {\bibinfo {volume} {2}},\ \bibinfo {eid} {033347} (\bibinfo {year}
  {2020})},\ \Eprint {https://arxiv.org/abs/1912.04292} {arXiv:1912.04292
  [cond-mat.stat-mech]} \BibitemShut {NoStop}%
\bibitem [{\citenamefont {Cubitt}(2023)}]{2023_cubitt}%
  \BibitemOpen
  \bibfield  {author} {\bibinfo {author} {\bibfnamefont {T.~S.}\ \bibnamefont
  {Cubitt}},\ }\href@noop {} {\  (\bibinfo {year} {2023})},\ \Eprint
  {https://arxiv.org/abs/2303.11962} {arXiv:2303.11962 [quant-ph]} \BibitemShut
  {NoStop}%
\bibitem [{\citenamefont {{Ding}}\ \emph {et~al.}(2024)\citenamefont {{Ding}},
  \citenamefont {{Chen}},\ and\ \citenamefont {{Lin}}}]{2024_Ding}%
  \BibitemOpen
  \bibfield  {author} {\bibinfo {author} {\bibfnamefont {Z.}~\bibnamefont
  {{Ding}}}, \bibinfo {author} {\bibfnamefont {C.-F.}\ \bibnamefont {{Chen}}},\
  and\ \bibinfo {author} {\bibfnamefont {L.}~\bibnamefont {{Lin}}},\ }\href
  {https://doi.org/10.1103/PhysRevResearch.6.033147} {\bibfield  {journal}
  {\bibinfo  {journal} {Physical Review Research}\ }\textbf {\bibinfo {volume}
  {6}},\ \bibinfo {eid} {033147} (\bibinfo {year} {2024})},\ \Eprint
  {https://arxiv.org/abs/2308.15676} {arXiv:2308.15676 [quant-ph]} \BibitemShut
  {NoStop}%
\bibitem [{\citenamefont {Li}\ \emph {et~al.}(2025)\citenamefont {Li},
  \citenamefont {Zhan},\ and\ \citenamefont {Lin}}]{Li_2024}%
  \BibitemOpen
  \bibfield  {author} {\bibinfo {author} {\bibfnamefont {H.-E.}\ \bibnamefont
  {Li}}, \bibinfo {author} {\bibfnamefont {Y.}~\bibnamefont {Zhan}},\ and\
  \bibinfo {author} {\bibfnamefont {L.}~\bibnamefont {Lin}},\ }\href
  {https://doi.org/10.1038/s41534-025-01124-8} {\bibfield  {journal} {\bibinfo
  {journal} {npj Quantum Inf.}\ }\textbf {\bibinfo {volume} {11}},\ \bibinfo
  {pages} {183} (\bibinfo {year} {2025})},\ \Eprint
  {https://arxiv.org/abs/2411.01470} {arXiv:2411.01470 [quant-ph]} \BibitemShut
  {NoStop}%
\bibitem [{\citenamefont {{Eder}}\ \emph {et~al.}(2025)\citenamefont {{Eder}},
  \citenamefont {{Fin{\v{z}}gar}}, \citenamefont {{Braun}},\ and\ \citenamefont
  {{Mendl}}}]{Eder_2025}%
  \BibitemOpen
  \bibfield  {author} {\bibinfo {author} {\bibfnamefont {P.~J.}\ \bibnamefont
  {{Eder}}}, \bibinfo {author} {\bibfnamefont {J.~R.}\ \bibnamefont
  {{Fin{\v{z}}gar}}}, \bibinfo {author} {\bibfnamefont {S.}~\bibnamefont
  {{Braun}}},\ and\ \bibinfo {author} {\bibfnamefont {C.~B.}\ \bibnamefont
  {{Mendl}}},\ }\href {https://doi.org/10.1103/PhysRevA.111.042430} {\bibfield
  {journal} {\bibinfo  {journal} {\pra}\ }\textbf {\bibinfo {volume} {111}},\
  \bibinfo {eid} {042430} (\bibinfo {year} {2025})},\ \Eprint
  {https://arxiv.org/abs/2407.11782} {arXiv:2407.11782 [quant-ph]} \BibitemShut
  {NoStop}%
\bibitem [{\citenamefont {{Zhan}}\ \emph {et~al.}(2026)\citenamefont {{Zhan}},
  \citenamefont {{Ding}}, \citenamefont {{Huhn}}, \citenamefont {{Gray}},
  \citenamefont {{Preskill}}, \citenamefont {{Chan}},\ and\ \citenamefont
  {{Lin}}}]{2026_Zhan}%
  \BibitemOpen
  \bibfield  {author} {\bibinfo {author} {\bibfnamefont {Y.}~\bibnamefont
  {{Zhan}}}, \bibinfo {author} {\bibfnamefont {Z.}~\bibnamefont {{Ding}}},
  \bibinfo {author} {\bibfnamefont {J.}~\bibnamefont {{Huhn}}}, \bibinfo
  {author} {\bibfnamefont {J.}~\bibnamefont {{Gray}}}, \bibinfo {author}
  {\bibfnamefont {J.}~\bibnamefont {{Preskill}}}, \bibinfo {author}
  {\bibfnamefont {G.~K.-L.}\ \bibnamefont {{Chan}}},\ and\ \bibinfo {author}
  {\bibfnamefont {L.}~\bibnamefont {{Lin}}},\ }\href
  {https://doi.org/10.1103/wzb3-dbg9} {\bibfield  {journal} {\bibinfo
  {journal} {Physical Review X}\ }\textbf {\bibinfo {volume} {16}},\ \bibinfo
  {eid} {011004} (\bibinfo {year} {2026})},\ \Eprint
  {https://arxiv.org/abs/2503.15827} {arXiv:2503.15827 [quant-ph]} \BibitemShut
  {NoStop}%
\bibitem [{\citenamefont {{Ding}}\ \emph {et~al.}(2025)\citenamefont {{Ding}},
  \citenamefont {{Zhan}}, \citenamefont {{Preskill}},\ and\ \citenamefont
  {{Lin}}}]{2026_Ding}%
  \BibitemOpen
  \bibfield  {author} {\bibinfo {author} {\bibfnamefont {Z.}~\bibnamefont
  {{Ding}}}, \bibinfo {author} {\bibfnamefont {Y.}~\bibnamefont {{Zhan}}},
  \bibinfo {author} {\bibfnamefont {J.}~\bibnamefont {{Preskill}}},\ and\
  \bibinfo {author} {\bibfnamefont {L.}~\bibnamefont {{Lin}}},\ }\href
  {https://doi.org/10.48550/arXiv.2508.05703} {\bibfield  {journal} {\bibinfo
  {journal} {arXiv e-prints}\ ,\ \bibinfo {eid} {arXiv:2508.05703}} (\bibinfo
  {year} {2025})},\ \Eprint {https://arxiv.org/abs/2508.05703}
  {arXiv:2508.05703 [quant-ph]} \BibitemShut {NoStop}%
\bibitem [{\citenamefont {{Mi}}\ \emph {et~al.}(2024)\citenamefont {{Mi}} \emph
  {et~al.}}]{2024_google_science}%
  \BibitemOpen
  \bibfield  {author} {\bibinfo {author} {\bibfnamefont {X.}~\bibnamefont
  {{Mi}}} \emph {et~al.},\ }\href {https://doi.org/10.1126/science.adh9932}
  {\bibfield  {journal} {\bibinfo  {journal} {Science}\ }\textbf {\bibinfo
  {volume} {383}},\ \bibinfo {pages} {1332} (\bibinfo {year} {2024})},\ \Eprint
  {https://arxiv.org/abs/2304.13878} {arXiv:2304.13878 [quant-ph]} \BibitemShut
  {NoStop}%
\bibitem [{\citenamefont {{Lloyd}}\ \emph {et~al.}(2024)\citenamefont
  {{Lloyd}}, \citenamefont {{Michailidis}}, \citenamefont {{Mi}}, \citenamefont
  {{Smelyanskiy}},\ and\ \citenamefont {{Abanin}}}]{2025_Lloyd}%
  \BibitemOpen
  \bibfield  {author} {\bibinfo {author} {\bibfnamefont {J.}~\bibnamefont
  {{Lloyd}}}, \bibinfo {author} {\bibfnamefont {A.}~\bibnamefont
  {{Michailidis}}}, \bibinfo {author} {\bibfnamefont {X.}~\bibnamefont {{Mi}}},
  \bibinfo {author} {\bibfnamefont {V.}~\bibnamefont {{Smelyanskiy}}},\ and\
  \bibinfo {author} {\bibfnamefont {D.~A.}\ \bibnamefont {{Abanin}}},\ }\href
  {https://doi.org/10.48550/arXiv.2404.12175} {\bibfield  {journal} {\bibinfo
  {journal} {arXiv e-prints}\ ,\ \bibinfo {eid} {arXiv:2404.12175}} (\bibinfo
  {year} {2024})},\ \Eprint {https://arxiv.org/abs/2404.12175}
  {arXiv:2404.12175 [quant-ph]} \BibitemShut {NoStop}%
\bibitem [{\citenamefont {{D{\"o}rstel}}\ \emph {et~al.}(2026)\citenamefont
  {{D{\"o}rstel}}, \citenamefont {{Iadecola}}, \citenamefont {{Wilson}},\ and\
  \citenamefont {{Buchhold}}}]{2026_buchhold_FF}%
  \BibitemOpen
  \bibfield  {author} {\bibinfo {author} {\bibfnamefont {T.}~\bibnamefont
  {{D{\"o}rstel}}}, \bibinfo {author} {\bibfnamefont {T.}~\bibnamefont
  {{Iadecola}}}, \bibinfo {author} {\bibfnamefont {J.~H.}\ \bibnamefont
  {{Wilson}}},\ and\ \bibinfo {author} {\bibfnamefont {M.}~\bibnamefont
  {{Buchhold}}},\ }\href {https://doi.org/10.1103/zp25-ppgk} {\bibfield
  {journal} {\bibinfo  {journal} {\prl}\ }\textbf {\bibinfo {volume} {136}},\
  \bibinfo {eid} {120603} (\bibinfo {year} {2026})},\ \Eprint
  {https://arxiv.org/abs/2510.24845} {arXiv:2510.24845 [quant-ph]} \BibitemShut
  {NoStop}%
\bibitem [{\citenamefont {{Feldmeier}}\ \emph {et~al.}(2026)\citenamefont
  {{Feldmeier}}, \citenamefont {{Liu}}, \citenamefont {{Lukin}},\ and\
  \citenamefont {{Choi}}}]{2026_choi_ff}%
  \BibitemOpen
  \bibfield  {author} {\bibinfo {author} {\bibfnamefont {J.}~\bibnamefont
  {{Feldmeier}}}, \bibinfo {author} {\bibfnamefont {Y.-J.}\ \bibnamefont
  {{Liu}}}, \bibinfo {author} {\bibfnamefont {M.~D.}\ \bibnamefont {{Lukin}}},\
  and\ \bibinfo {author} {\bibfnamefont {S.}~\bibnamefont {{Choi}}},\ }\href
  {https://doi.org/10.48550/arXiv.2603.10119} {\bibfield  {journal} {\bibinfo
  {journal} {arXiv e-prints}\ ,\ \bibinfo {eid} {arXiv:2603.10119}} (\bibinfo
  {year} {2026})},\ \Eprint {https://arxiv.org/abs/2603.10119}
  {arXiv:2603.10119 [quant-ph]} \BibitemShut {NoStop}%
\bibitem [{\citenamefont {{Langbehn}}\ \emph {et~al.}(2024)\citenamefont
  {{Langbehn}}, \citenamefont {{Snizhko}}, \citenamefont {{Gornyi}},
  \citenamefont {{Morigi}}, \citenamefont {{Gefen}},\ and\ \citenamefont
  {{Koch}}}]{2024_langbehn}%
  \BibitemOpen
  \bibfield  {author} {\bibinfo {author} {\bibfnamefont {J.}~\bibnamefont
  {{Langbehn}}}, \bibinfo {author} {\bibfnamefont {K.}~\bibnamefont
  {{Snizhko}}}, \bibinfo {author} {\bibfnamefont {I.}~\bibnamefont {{Gornyi}}},
  \bibinfo {author} {\bibfnamefont {G.}~\bibnamefont {{Morigi}}}, \bibinfo
  {author} {\bibfnamefont {Y.}~\bibnamefont {{Gefen}}},\ and\ \bibinfo {author}
  {\bibfnamefont {C.~P.}\ \bibnamefont {{Koch}}},\ }\href
  {https://doi.org/10.1103/PRXQuantum.5.030301} {\bibfield  {journal} {\bibinfo
   {journal} {PRX Quantum}\ }\textbf {\bibinfo {volume} {5}},\ \bibinfo {eid}
  {030301} (\bibinfo {year} {2024})},\ \Eprint
  {https://arxiv.org/abs/2311.05258} {arXiv:2311.05258 [quant-ph]} \BibitemShut
  {NoStop}%
\bibitem [{\citenamefont {{Bhuiyan}}\ \emph {et~al.}(2026)\citenamefont
  {{Bhuiyan}}, \citenamefont {{Pan}},\ and\ \citenamefont
  {{Jian}}}]{2026_jian_adaptive}%
  \BibitemOpen
  \bibfield  {author} {\bibinfo {author} {\bibfnamefont {A.}~\bibnamefont
  {{Bhuiyan}}}, \bibinfo {author} {\bibfnamefont {H.}~\bibnamefont {{Pan}}},\
  and\ \bibinfo {author} {\bibfnamefont {C.-M.}\ \bibnamefont {{Jian}}},\
  }\href {https://doi.org/10.1103/nk38-ygyq} {\bibfield  {journal} {\bibinfo
  {journal} {Physical Review Research}\ }\textbf {\bibinfo {volume} {8}},\
  \bibinfo {eid} {023147} (\bibinfo {year} {2026})},\ \Eprint
  {https://arxiv.org/abs/2507.13437} {arXiv:2507.13437 [quant-ph]} \BibitemShut
  {NoStop}%
\bibitem [{\citenamefont {{Lu}}\ \emph {et~al.}(2026)\citenamefont {{Lu}},
  \citenamefont {{Liu}}, \citenamefont {{Gopalakrishnan}},\ and\ \citenamefont
  {{You}}}]{You_2026}%
  \BibitemOpen
  \bibfield  {author} {\bibinfo {author} {\bibfnamefont {T.-C.}\ \bibnamefont
  {{Lu}}}, \bibinfo {author} {\bibfnamefont {Y.-J.}\ \bibnamefont {{Liu}}},
  \bibinfo {author} {\bibfnamefont {S.}~\bibnamefont {{Gopalakrishnan}}},\ and\
  \bibinfo {author} {\bibfnamefont {Y.}~\bibnamefont {{You}}},\ }\href
  {https://doi.org/10.1103/lrqp-pnl1} {\bibfield  {journal} {\bibinfo
  {journal} {PRX Quantum}\ }\textbf {\bibinfo {volume} {7}},\ \bibinfo {eid}
  {033041} (\bibinfo {year} {2026})},\ \Eprint
  {https://arxiv.org/abs/2511.19597} {arXiv:2511.19597 [quant-ph]} \BibitemShut
  {NoStop}%
\bibitem [{\citenamefont {{Watts}}\ \emph {et~al.}(2026)\citenamefont
  {{Watts}}, \citenamefont {{Sarkar}}, \citenamefont {{Collins}}, \citenamefont
  {{Nguyen}}, \citenamefont {{Quezada}}, \citenamefont {{Bremner}},\ and\
  \citenamefont {{Elman}}}]{2026_elman_chem_TS}%
  \BibitemOpen
  \bibfield  {author} {\bibinfo {author} {\bibfnamefont {T.~W.}\ \bibnamefont
  {{Watts}}}, \bibinfo {author} {\bibfnamefont {S.}~\bibnamefont {{Sarkar}}},
  \bibinfo {author} {\bibfnamefont {D.}~\bibnamefont {{Collins}}}, \bibinfo
  {author} {\bibfnamefont {N.}~\bibnamefont {{Nguyen}}}, \bibinfo {author}
  {\bibfnamefont {L.}~\bibnamefont {{Quezada}}}, \bibinfo {author}
  {\bibfnamefont {M.~J.}\ \bibnamefont {{Bremner}}},\ and\ \bibinfo {author}
  {\bibfnamefont {S.~J.}\ \bibnamefont {{Elman}}},\ }\href
  {https://doi.org/10.1103/8s88-x43b} {\bibfield  {journal} {\bibinfo
  {journal} {\pra}\ }\textbf {\bibinfo {volume} {114}},\ \bibinfo {eid}
  {022471} (\bibinfo {year} {2026})},\ \Eprint
  {https://arxiv.org/abs/2602.11603} {arXiv:2602.11603 [quant-ph]} \BibitemShut
  {NoStop}%
\bibitem [{\citenamefont {{Zhou}}\ and\ \citenamefont {{Wu}}(2026)}]{2026_wu}%
  \BibitemOpen
  \bibfield  {author} {\bibinfo {author} {\bibfnamefont {Y.}~\bibnamefont
  {{Zhou}}}\ and\ \bibinfo {author} {\bibfnamefont {B.}~\bibnamefont {{Wu}}},\
  }\href {https://doi.org/10.48550/arXiv.2606.05815} {\bibfield  {journal}
  {\bibinfo  {journal} {arXiv e-prints}\ ,\ \bibinfo {eid} {arXiv:2606.05815}}
  (\bibinfo {year} {2026})},\ \Eprint {https://arxiv.org/abs/2606.05815}
  {arXiv:2606.05815 [quant-ph]} \BibitemShut {NoStop}%
\bibitem [{\citenamefont {Tang}\ \emph {et~al.}(2026)\citenamefont {Tang},
  \citenamefont {Barad},\ and\ \citenamefont {Wen}}]{2026_Tang}%
  \BibitemOpen
  \bibfield  {author} {\bibinfo {author} {\bibfnamefont {Q.}~\bibnamefont
  {Tang}}, \bibinfo {author} {\bibfnamefont {R.}~\bibnamefont {Barad}},\ and\
  \bibinfo {author} {\bibfnamefont {X.}~\bibnamefont {Wen}},\ }\href@noop {} {\
   (\bibinfo {year} {2026})},\ \Eprint {https://arxiv.org/abs/2606.19465}
  {arXiv:2606.19465 [cond-mat.stat-mech]} \BibitemShut {NoStop}%
\bibitem [{\citenamefont {{Bai}}(2026)}]{2026_Bai}%
  \BibitemOpen
  \bibfield  {author} {\bibinfo {author} {\bibfnamefont {C.}~\bibnamefont
  {{Bai}}},\ }\href {https://doi.org/10.48550/arXiv.2607.08827} {\bibfield
  {journal} {\bibinfo  {journal} {arXiv e-prints}\ ,\ \bibinfo {eid}
  {arXiv:2607.08827}} (\bibinfo {year} {2026})},\ \Eprint
  {https://arxiv.org/abs/2607.08827} {arXiv:2607.08827 [cond-mat.stat-mech]}
  \BibitemShut {NoStop}%
\bibitem [{\citenamefont {Fang}\ \emph {et~al.}(2025)\citenamefont {Fang},
  \citenamefont {Lu},\ and\ \citenamefont {Tong}}]{Fang_2024}%
  \BibitemOpen
  \bibfield  {author} {\bibinfo {author} {\bibfnamefont {D.}~\bibnamefont
  {Fang}}, \bibinfo {author} {\bibfnamefont {J.}~\bibnamefont {Lu}},\ and\
  \bibinfo {author} {\bibfnamefont {Y.}~\bibnamefont {Tong}},\ }\href
  {https://doi.org/10.1103/PhysRevLett.134.140405} {\bibfield  {journal}
  {\bibinfo  {journal} {Phys. Rev. Lett.}\ }\textbf {\bibinfo {volume} {134}},\
  \bibinfo {pages} {140405} (\bibinfo {year} {2025})},\ \Eprint
  {https://arxiv.org/abs/2404.11503} {arXiv:2404.11503 [quant-ph]} \BibitemShut
  {NoStop}%
\bibitem [{\citenamefont {{Lin}}(2025)}]{Lin_2025}%
  \BibitemOpen
  \bibfield  {author} {\bibinfo {author} {\bibfnamefont {L.}~\bibnamefont
  {{Lin}}},\ }\href {https://doi.org/10.48550/arXiv.2505.21308} {\bibfield
  {journal} {\bibinfo  {journal} {arXiv e-prints}\ ,\ \bibinfo {eid}
  {arXiv:2505.21308}} (\bibinfo {year} {2025})},\ \Eprint
  {https://arxiv.org/abs/2505.21308} {arXiv:2505.21308 [quant-ph]} \BibitemShut
  {NoStop}%
\bibitem [{\citenamefont {{Seki}}\ \emph {et~al.}(2026)\citenamefont {{Seki}},
  \citenamefont {{Kikuchi}}, \citenamefont {{Hayata}},\ and\ \citenamefont
  {{Yunoki}}}]{Seki_2026}%
  \BibitemOpen
  \bibfield  {author} {\bibinfo {author} {\bibfnamefont {K.}~\bibnamefont
  {{Seki}}}, \bibinfo {author} {\bibfnamefont {Y.}~\bibnamefont {{Kikuchi}}},
  \bibinfo {author} {\bibfnamefont {T.}~\bibnamefont {{Hayata}}},\ and\
  \bibinfo {author} {\bibfnamefont {S.}~\bibnamefont {{Yunoki}}},\ }\href
  {https://doi.org/10.1103/rjdv-bxgq} {\bibfield  {journal} {\bibinfo
  {journal} {Physical Review Applied}\ }\textbf {\bibinfo {volume} {26}},\
  \bibinfo {eid} {014062} (\bibinfo {year} {2026})},\ \Eprint
  {https://arxiv.org/abs/2601.08137} {arXiv:2601.08137 [quant-ph]} \BibitemShut
  {NoStop}%
\bibitem [{\citenamefont {Dengis}\ \emph {et~al.}(2014)\citenamefont {Dengis},
  \citenamefont {Koenig},\ and\ \citenamefont {Pastawski}}]{Dengis_2014}%
  \BibitemOpen
  \bibfield  {author} {\bibinfo {author} {\bibfnamefont {J.}~\bibnamefont
  {Dengis}}, \bibinfo {author} {\bibfnamefont {R.}~\bibnamefont {Koenig}},\
  and\ \bibinfo {author} {\bibfnamefont {F.}~\bibnamefont {Pastawski}},\ }\href
  {https://doi.org/10.1088/1367-2630/16/1/013023} {\bibfield  {journal}
  {\bibinfo  {journal} {New J. Phys.}\ }\textbf {\bibinfo {volume} {16}},\
  \bibinfo {pages} {013023} (\bibinfo {year} {2014})},\ \Eprint
  {https://arxiv.org/abs/1310.1036} {arXiv:1310.1036 [quant-ph]} \BibitemShut
  {NoStop}%
\bibitem [{\citenamefont {Koenig}\ and\ \citenamefont
  {Pastawski}(2014)}]{Koenig_2013}%
  \BibitemOpen
  \bibfield  {author} {\bibinfo {author} {\bibfnamefont {R.}~\bibnamefont
  {Koenig}}\ and\ \bibinfo {author} {\bibfnamefont {F.}~\bibnamefont
  {Pastawski}},\ }\href {https://doi.org/10.1103/PhysRevB.90.045101} {\bibfield
   {journal} {\bibinfo  {journal} {Phys. Rev. B}\ }\textbf {\bibinfo {volume}
  {90}},\ \bibinfo {pages} {045101} (\bibinfo {year} {2014})},\ \Eprint
  {https://arxiv.org/abs/1310.1037} {arXiv:1310.1037 [quant-ph]} \BibitemShut
  {NoStop}%
\bibitem [{\citenamefont {{Liu}}\ \emph {et~al.}(2021)\citenamefont {{Liu}},
  \citenamefont {{Bergholtz}},\ and\ \citenamefont {{Budich}}}]{Liu_2021}%
  \BibitemOpen
  \bibfield  {author} {\bibinfo {author} {\bibfnamefont {Z.}~\bibnamefont
  {{Liu}}}, \bibinfo {author} {\bibfnamefont {E.~J.}\ \bibnamefont
  {{Bergholtz}}},\ and\ \bibinfo {author} {\bibfnamefont {J.~C.}\ \bibnamefont
  {{Budich}}},\ }\href {https://doi.org/10.1103/PhysRevResearch.3.043119}
  {\bibfield  {journal} {\bibinfo  {journal} {Physical Review Research}\
  }\textbf {\bibinfo {volume} {3}},\ \bibinfo {eid} {043119} (\bibinfo {year}
  {2021})},\ \Eprint {https://arxiv.org/abs/2108.10327} {arXiv:2108.10327
  [cond-mat.str-el]} \BibitemShut {NoStop}%
\bibitem [{\citenamefont {{Elman}}(2026)}]{elman_2026_local_filter}%
  \BibitemOpen
  \bibfield  {author} {\bibinfo {author} {\bibfnamefont {S.~J.}\ \bibnamefont
  {{Elman}}},\ }\href {https://doi.org/10.48550/arXiv.2609.31266} {\bibfield
  {journal} {\bibinfo  {journal} {arXiv e-prints}\ ,\ \bibinfo {eid}
  {arXiv:2609.31266}} (\bibinfo {year} {2026})},\ \Eprint
  {https://arxiv.org/abs/2609.31266} {arXiv:2609.31266 [quant-ph]} \BibitemShut
  {NoStop}%
\bibitem [{Note1()}]{Note1}%
  \BibitemOpen
  \bibinfo {note} {It is in general not a projection, since every channel can
  couple to each mode with different decay rates.}\BibitemShut {Stop}%
\bibitem [{Note2()}]{Note2}%
  \BibitemOpen
  \bibinfo {note} {Here $h(q)$ is of rank $2s$, so that any power higher than
  $2s-1$ can be expressed in terms of lower powers $k=0,1,\protect \dots
  ,2s-1$.}\BibitemShut {Stop}%
\bibitem [{Note3()}]{Note3}%
  \BibitemOpen
  \bibinfo {note} {The existence of such choice is ensured by the Rellich's
  theorem~\cite {Rellich1937, Kato1995}. It states that along a real-analytic
  path $q(\lambda )$ in the BZ, the Hermitian matrix $h[q(\lambda )]$ (which is
  analytic in $q$ followed by the real-space locality of $h$) allows to choose
  its eigenvalues $\varepsilon _n(\lambda )$ and corresponding eigenvectors
  $u_n(\lambda )$ as analytic functions of $\lambda $.}\BibitemShut {Stop}%
\bibitem [{\citenamefont {{Zhang}}\ and\ \citenamefont
  {{Barthel}}(2022)}]{2022_Zhang}%
  \BibitemOpen
  \bibfield  {author} {\bibinfo {author} {\bibfnamefont {Y.}~\bibnamefont
  {{Zhang}}}\ and\ \bibinfo {author} {\bibfnamefont {T.}~\bibnamefont
  {{Barthel}}},\ }\href {https://doi.org/10.1103/PhysRevLett.129.120401}
  {\bibfield  {journal} {\bibinfo  {journal} {\prl}\ }\textbf {\bibinfo
  {volume} {129}},\ \bibinfo {eid} {120401} (\bibinfo {year} {2022})},\ \Eprint
  {https://arxiv.org/abs/2204.05346} {arXiv:2204.05346 [quant-ph]} \BibitemShut
  {NoStop}%
\bibitem [{\citenamefont {{Westhoff}}\ \emph {et~al.}(2026)\citenamefont
  {{Westhoff}}, \citenamefont {{Murciano}},\ and\ \citenamefont
  {{Moroder}}}]{2026_Westhoff}%
  \BibitemOpen
  \bibfield  {author} {\bibinfo {author} {\bibfnamefont {P.}~\bibnamefont
  {{Westhoff}}}, \bibinfo {author} {\bibfnamefont {S.}~\bibnamefont
  {{Murciano}}},\ and\ \bibinfo {author} {\bibfnamefont {M.}~\bibnamefont
  {{Moroder}}},\ }\href {https://doi.org/10.48550/arXiv.2610.13784} {\bibfield
  {journal} {\bibinfo  {journal} {arXiv e-prints}\ ,\ \bibinfo {eid}
  {arXiv:2610.13784}} (\bibinfo {year} {2026})},\ \Eprint
  {https://arxiv.org/abs/2610.13784} {arXiv:2610.13784 [quant-ph]} \BibitemShut
  {NoStop}%
\bibitem [{\citenamefont {Rellich}(1937)}]{Rellich1937}%
  \BibitemOpen
  \bibfield  {author} {\bibinfo {author} {\bibfnamefont {F.}~\bibnamefont
  {Rellich}},\ }\href {https://doi.org/10.1007/BF01571652} {\bibfield
  {journal} {\bibinfo  {journal} {Mathematische Annalen}\ }\textbf {\bibinfo
  {volume} {113}},\ \bibinfo {pages} {600} (\bibinfo {year}
  {1937})}\BibitemShut {NoStop}%
\bibitem [{\citenamefont {Kato}(1995)}]{Kato1995}%
  \BibitemOpen
  \bibfield  {author} {\bibinfo {author} {\bibfnamefont {T.}~\bibnamefont
  {Kato}},\ }\href {https://doi.org/10.1007/978-3-642-66282-9} {\emph {\bibinfo
  {title} {Perturbation Theory for Linear Operators}}},\ \bibinfo {edition}
  {2nd}\ ed.,\ Classics in Mathematics\ (\bibinfo  {publisher} {Springer},\
  \bibinfo {address} {Berlin, Heidelberg},\ \bibinfo {year} {1995})\BibitemShut
  {NoStop}%
\end{thebibliography}%

\appendix

\subsection{Spectral-filtering realization and mixing time for short-range correlations}

The local jump construction in Eq.~\eqref{eq:linear-weight-jump-coefficient} admits a spectral-filtering realization \cite{2024_Ding,2026_Zhan} that can be specified without detailed prior knowledge about the spectral information of the target Hamiltonian. 
Jump operators $\{ K_\mu \}$ for ground-state preparation can be constructed from a set of local coupling operators $\{ A_\mu \}$. 
Resolving each $A_\mu$ in the energy eigenbasis $\mathbf H |\Phi_i\rangle = \mathcal E_i |\Phi_i\rangle$ and applying a filter $f(\omega)$ gives 
\be
K_\mu
=
\sum_{i,j}
f(\mathcal E_i-\mathcal E_j)
|\Phi_i\rangle\langle\Phi_i|
A_\mu
|\Phi_j\rangle\langle\Phi_j|.
\label{eq:jump-filter-eigenbasis}
\ee
Choosing $f(\omega)=0$ for $\omega\geq0$ removes all energy-preserving and energy-raising transitions and ensures the dark-state condition in Eq.~\eqref{eq:dark-state-condition}.

For free-fermion target Hamiltonians, one takes the complete set of on-site Nambu operators, $A_\mu=\Psi_\mu$ with $\mu=(j,a)$, as the coupling operators, and set
\be
f(\omega)=
\begin{cases}
\sqrt{\kappa}\,\omega,
    &-E_{\max}\leq\omega<0,\\
0,&\omega\geq0,
\end{cases}
\label{eq:linear-scalar-filter}
\ee
where $\kappa>0$ determines the jump rate scale. 
This choice realizes our construction in Eq.~\eqref{eq:linear-weight-jump-coefficient}, with $G=f(-h)=\sqrt{\kappa}\,hP_+$. 
Since the spectral bound $E_{\max} \ge \|h\|$ is a $\mathcal{O}(1)$ constant, a smooth continuation of $f(\omega)$ outside of the target energy window $\omega \notin [-E_{\rm max}, E_{\rm max}]$ can be chosen with a finite cutoff scale. 
This continuation does not affect the jump operators, and is important for a practical realization as we discuss below. 

In particular, while Eq.~\eqref{eq:jump-filter-eigenbasis} provides a concrete construction of jumps by spectral filtering, its direct evaluation would require explicit knowledge of the eigenvalues and eigenstates of $\mathbf H$. 
Instead, one consider a Fourier transformation that translates the jump operator into a time-domain integral representation
\be
K_\mu
=
\int_{\mathbb R}\widetilde f(\tau)A_\mu(\tau)\,\mathrm dt
=
\int_{\mathbb R}
\widetilde f(\tau)
e^{i\mathbf H\tau}A_\mu e^{-i\mathbf H\tau}
\,\mathrm d\tau ,
\label{eq:jump-filter-time-integral}
\ee
with $\widetilde f(\tau)=(2\pi)^{-1}\int_{\mathbb R}f(\omega)e^{-i\omega \tau}\,\mathrm d\omega$. 
Truncating the integral to $|\tau|\leq T$ induces an operator-norm error bounded by $\|A_\mu\|\int_{|\tau|>T}|\widetilde f(t)|\,\mathrm dt$. 
For the filter in Eq.~\eqref{eq:linear-scalar-filter}, the derivative discontinuity at zero and the smooth continuation give
\be
|\widetilde f(\tau)|
=
\mathcal O\!\left(
\frac{\sqrt{\kappa}}
     {(1+|\tau|)^2}
\right).
\label{eq:linear-filter-time-decay}
\ee
Since $\|A_\mu\|=1$, the truncation error is $\mathcal O[\sqrt{\kappa}/(1+T)]$. At a fixed $\kappa$, a simulation $T=\mathcal O(\eta^{-1})$ is therefore sufficient for reaching an operator-norm error within $\eta$ through Eq.~\eqref{eq:jump-filter-time-integral}, leading to an efficient evaluation of the jump operators \cite{2024_Ding,2026_Zhan}. 

Nevertheless, efficient evaluation of the jumps does not by itself imply rapid ground-state preparation by using them.
For the filter in Eq.~\eqref{eq:linear-scalar-filter} and a complete set of coupling operator $A_\mu=\Psi_\mu$ with $\mu=(j,a)$, the positive-energy modes ($\varepsilon_\nu > 0$) are subjected to independent losses with rates $\Gamma_\nu = \kappa\varepsilon_\nu^2$,
with $\langle\gamma_\nu^\dagger\gamma_\nu\rangle_t =e^{-\Gamma_\nu t} \langle\gamma_\nu^\dagger\gamma_\nu\rangle_0$. 
Let $\Delta_h=\min_{\varepsilon_\nu>0}\varepsilon_\nu$, one has 
\be
\begin{aligned}
\operatorname{Tr}[(\Id-P_{\rm GS})\rho(t)]
&\leq
\sum_{\nu:\varepsilon_\nu>0}
\langle\gamma_\nu^\dagger\gamma_\nu\rangle_t\\
&\quad\leq
N e^{-\kappa\Delta_h^2t}.
\end{aligned}
\ee
Here for the second inequality, the identity $\Id - P_{\rm GS} < \sum_{\nu: \varepsilon_\nu>0} \gamma_\nu^\dagger \gamma_\nu$ is used. 

To quantify preparation of a possibly degenerate ground-state subspace, define the trace distance
\be
d_{\rm GS}(\rho)
=
\inf_{\sigma=P_{\rm GS}\sigma P_{\rm GS}}
\frac12\|\rho-\sigma\|_1,
\label{eq:ground-space-distance}
\ee
where the infimum is over density matrices and $\|\cdot\|_1$ denotes the trace norm.
This distance is upper bounded by $d_{\rm GS}(\rho) \le \sqrt{\operatorname{Tr}[(\Id-P_{\rm GS})\rho]}$.
The mixing time $\tau_{\rm mix}(\epsilon)$, defined as the smallest time at which $d_{\rm GS}(\rho(t))\leq\epsilon$ for every initial state, is then given by
\be
\tau_{\rm mix}(\epsilon)
\leq
\frac{1}{\kappa\Delta_h^2}
\log\frac{N}{\epsilon^2},
\qquad 0<\epsilon<1.
\ee
Together with Eq.~\eqref{eq:linear-filter-time-decay}, the filter in in Eq.~\eqref{eq:linear-scalar-filter} allows an efficient evaluation of a set of exponentially local jump operators for rapid ground-state preparation for ground states with short-range correlations. 

For comparison, the uniform-weight choice, which is adopted in Ref.~\cite{2026_Zhan} for realizing rapid ground-state preparation, requires
\be
f(-\varepsilon_\nu)=\sqrt{\kappa}
\quad\text{for }\varepsilon_\nu>0,
\qquad
f(\omega)=0\quad\text{for }\omega\geq0 .
\label{eq:uniform-filter-values}
\ee
For the chosen on-site coupling operators $A_\mu = \Psi_\mu$ with $\mu=(j,a)$, such a filter gives $G=\sqrt{\kappa}P_+$, which is a natural choice for the ground-state preparation. 
In particular, to evaluate the corresponding jumps through the time-domain representation in Eq.~\eqref{eq:jump-filter-time-integral}, one can reproduce these sampled values $\omega = \varepsilon_\nu > 0$ with a smooth filter $f$ by choosing a smooth continuation in the gap regime $\omega \in (0, \Delta_h)$ together with a smooth cutoff about the spectral bound $E_{\max}$.
Specifying such a smooth continuation generally requires knowing a positive lower bound on $\Delta_h$.

Moreover, under the uniform-weight setting, every positive-energy mode decays at the same rate $\kappa$, which leads to 
\be
\tau_{\rm mix}^{\rm uniform}(\epsilon)
\leq
\frac{1}{\kappa}\log\frac{N}{\epsilon^2}.
\label{eq:uniform-mixing}
\ee
When $\Delta_h$ has a size-independent positive lower bound, i.e. when the target Hamiltonian is gapped, both the uniform weights in Eq.~\eqref{eq:uniform-filter-values} and linear weights in Eq.~\eqref{eq:linear-scalar-filter} give a bound on mixing time that is logarithmic in the system size $N$ for a fixed accuracy.
Although the linear filter contains an additional factor $(1/\Delta_h)^2$ in the bound on mixing time, it is a constant for gapped systems. 
The liner weights nevertheless avoid a gap-dependent smooth continuation for the filter that is required by an efficient evaluation of the jump operators through Eq.~\eqref{eq:jump-filter-time-integral}.

\end{document}